\PassOptionsToPackage{unicode}{hyperref}
\PassOptionsToPackage{hyphens}{url}
\documentclass{article}

\usepackage{arxiv}

\usepackage{amsmath,amssymb}
\usepackage{iftex}
\usepackage[T1]{fontenc}
\usepackage[utf8]{inputenc}
\usepackage{textcomp} 
\usepackage{lmodern}
\usepackage{xcolor}
\usepackage{graphicx}
\usepackage{natbib}

\title{A proposal for homoscedastic modelling with conditional auto-regressive distributions}
\author{Miguel A. Martinez-Beneito\\  Department of Statistics and Operations Research, University of Valencia, Valencia, Spain \And
Aritz Adin\\ Department of Statistics, Computer Science and Mathematics, Public University of Navarre, Pamplona, Spain\\ Institute of Advanced Materials and Mathematics (InaMat2), Public University of Navarre, Pamplona, Spain \And
Tomás Goicoa\\ Department of Statistics, Computer Science and Mathematics, Public University of Navarre, Pamplona, Spain\\ Institute of Advanced Materials and Mathematics (InaMat2), Public University of Navarre, Pamplona, Spain \\ Research Network on Health Services in Chronic Diseases (REDISSEC), Madrid, Spain \And
Lola Ugarte\\ Department of Statistics, Computer Science and Mathematics, Public University of Navarre, Pamplona, Spain\\ Institute of Advanced Materials and Mathematics (InaMat2), Public University of Navarre, Pamplona, Spain \\ Department of Mathematics, UNED, Pamplona, Spain}

\begin{document}

\maketitle

\begin{abstract}
Conditional auto-regressive (CAR) distributions are widely used to induce spatial dependence in the geographic analysis of areal data. These distributions establish multivariate dependence networks by defining conditional relationships between neighboring units, resulting in positive dependence among nearby observations. Despite their practical convenience, the conditional nature of CAR distributions can lead to undesirable marginal properties, such as inherent heterogeneity assumptions that may significantly impact the posterior distributions.

In this paper, we highlight the variance issues associated with CAR distributions, particularly focusing on edge effects and artifacts related to the region's geometry. We show that edge effects may be more significant and widespread in the outcomes of disease mapping studies than previously anticipated. To address these homoscedasticity concerns, we introduce a new conditional autoregressive distribution designed to mitigate these problems. We demonstrate how this distribution effectively resolves the practical issues identified in earlier models.
\end{abstract}

\section{Introduction}

Disease mapping has been a highly active area of research over the past three decades, with numerous methodological advancements and practical applications constantly published in the field of epidemiology. The primary objective of disease mapping is to address the small area estimation challenges inherent in public health analyses. Epidemiologists aim to analyze study units as small as possible to ensure that underlying spatial risks are not obscured by aggregation across larger areas. However, smaller units typically have fewer individuals, which can compromise the reliability of risk estimates, particularly for rare diseases.

Moreover, the administrative areal units used in these studies often differ significantly in population size and composition, adding complexity to making accurate inferences about the spatial distribution of risks. If risks were treated as independent across different areal units, their variances would be inversely proportional to the number of expected cases (\cite{Martinez-Beneito.BotellaRocamora2019}, page 112). As a result, risk estimates for sparsely populated areas tend to be more variable and often exhibit extreme values, irrespective of the actual risks. The primary goal of disease mapping is to stabilize these risk estimates in less populated areas, mitigating the mentioned artifacts and providing more accurate representations of the underlying risks.

Although early models in disease mapping did not account for spatial dependence between areal units \citep{Clayton.Kaldor1987}, spatial dependence soon emerged as a crucial tool for addressing the small area estimation issues inherent in disease mapping studies. Among the various approaches, conditional auto-regressive (CAR) distributions, specifically designed for areal data, have become the most widely used method for inducing spatial dependence in these problems.

CAR distributions can be formulated either multivariately, defining the joint multivariate distribution \(P(\pmb{\theta})\) for a random vector \(\pmb{\theta}\) (commonly referred to as \emph{Gaussian Markov Random Fields} or \emph{GMRF}), or conditionally, through the full conditionals \(P(\theta_i\mid \pmb{\theta}_{-i})\) for each unit $i$. In the first case, \(\pmb{\theta}\) follows a multivariate Normal distribution with a given precision matrix \(\tau \mathbf{Q}\). In the second, the conditional distributions are a set of univariate normal distributions, where the means depend on neighbouring locations. Both approaches explicitly model the spatial relationships between units and their neighbors. In the multivariate case, these relationships are encapsulated in the precision matrix, which defines the conditional dependencies between neighboring units.

Continuous Gaussian fields have also been applied to disease mapping problems \citep{Kelsall.Wakefield2002, Berke2004, Best.Richardson.ea2005}. Although Gaussian fields are not inherently suited for areal data analysis due to their continuous spatial nature, they can be adapted for this purpose. These models assume that the covariance between spatial units is a parametric function of their distance, often measured as the distance between centroids in the case of areal data. Unlike GMRFs, which model spatial dependence using precision matrices, Gaussian fields induce spatial dependence through the covariance matrix. Since covariance and precision matrices are mathematical inverses of each other,  which is a counter-intuitive mathematical transformation, the dependence structure defined with any of these approaches may not make full sense from the other point of view. Consequently, some properties of CAR-based random effects, designed with a conditional variance perspective, may exhibit less desirable behavior when viewed in terms of marginal variances. Similarly, Gaussian fields, optimized for marginal covariance, may not perform as well when evaluated from a conditional variance perspective.

The edge effect is a well-documented pathological feature of CAR-distributed random vectors \citep{Stern.Cressie1999, Lawson.Biggeri.ea1999a}. This effect causes the values of $\pmb{\theta}$ at the borders of the study region to exhibit greater variability than those within the interior. This variability introduces artifacts in spatial risk smoothing models, similar to those observed in low-population areas, potentially leading border regions to display extreme risk levels regardless of their actual risk. This is clearly an undesirable feature of spatial smoothing models that would be convenient to minimize for practical applications. Several approaches have been developed to address the edge effect, including weighting schemes for boundary areas to account for the degree of missing information at each location \citep{Dreassi.Biggeri1998} and the construction of external buffer zones with either missing observed data \citep{VidalRodeiro.Lawson2005} or real observed data when available \citep{OrozcoAcosta.Adin.ea2021, OrozcoAcosta.Adin.ea2023}. Despite significant attention to the edge effect in the 1990s and early 2000s, methodological proposals have been scarce over the past two decades, with much less focus on this issue in recent years.

The aim of this paper is to evaluate certain features of CAR-distributed random vectors from a marginal perspective. Based on this evaluation, we propose a modification of the traditional Intrinsic CAR distribution to address some identified issues, particularly those related to edge effects and additional features linked to the geometry of the graph representing the geographic structure of the study region.

The paper is structured as follows: In Section 2, we examine CAR-distributed random effects from a marginal perspective across various real-world contexts, specifically mortality studies in different Spanish regions. Section 3 introduces a methodological modification to the ICAR distribution designed to resolve the problematic features identified in Section 2. In Section 4, we assess the proposed modification through both simulated scenarios and the same real-world contexts discussed earlier. Finally, Section 5 concludes with a discussion of the findings and their implications.

\section{An empirical assessment of the marginal variances of the Intrinsic CAR distribution}

We now study some properties of the marginal variances of the Intrinsic CAR (ICAR) distribution in several real contexts. From a full-conditional point of view, we say that a random vector \(\pmb{\theta}\) follows an ICAR distribution if each of their components, given the rest, follow the univariate distribution
\[\theta_i\mid\pmb{\theta}_{-i}\sim N(\bar{\theta}_i,\tau \sum_{\{j:j\sim i\}} w_{ij}),\;i=1,\ldots,I,\]
where \(\bar{\theta}_i=(\sum_{\{j:j\sim i\}} w_{ij} \theta_j)/(\sum_{\{j:j\sim i\}} w_{ij})\) is a weighted mean of the values of \(\theta\) in the neighbouring areas of unit \(i\). In this expression the weights \(w_{ij}\) are intended to weight the influence of the neighbours \(j\) of each areal unit \(i\), which we denote as \(j\sim i\). In the previous expression, and from now on, the variability in normal distributions will be parameterized as a function of precisions, or precision matrices in the multivariate case. Alternatively, from a multivariate perspective, the ICAR distribution may be also defined as:
\[\pmb{\theta}\sim N_I(\mathbf{0}_I,\tau\mathbf{Q})\]
where \(\mathbf{Q}=\mathbf{D}-\mathbf{W}\) for the diagonal matrix \(\mathbf{D}\), with \(D_{ii}=\sum_{\{j:j\sim i\}}w_{ij}\), and a sparse weight matrix with \(W_{ij}=w_{ij}\) if \(i\sim j\), and 0 otherwise. For the rest of this section we will assume the typical assumption of adjacency for the spatial structure so that \(w_{ij}=1\) if \(i\sim j\) and 0 otherwise. We will refer to $\mathbf{Q}$, the precision matrix without the precision parameter $\tau$, as the structure matrix of the ICAR distribution.

\begin{figure}
\centering
\includegraphics[width=0.7\textwidth]{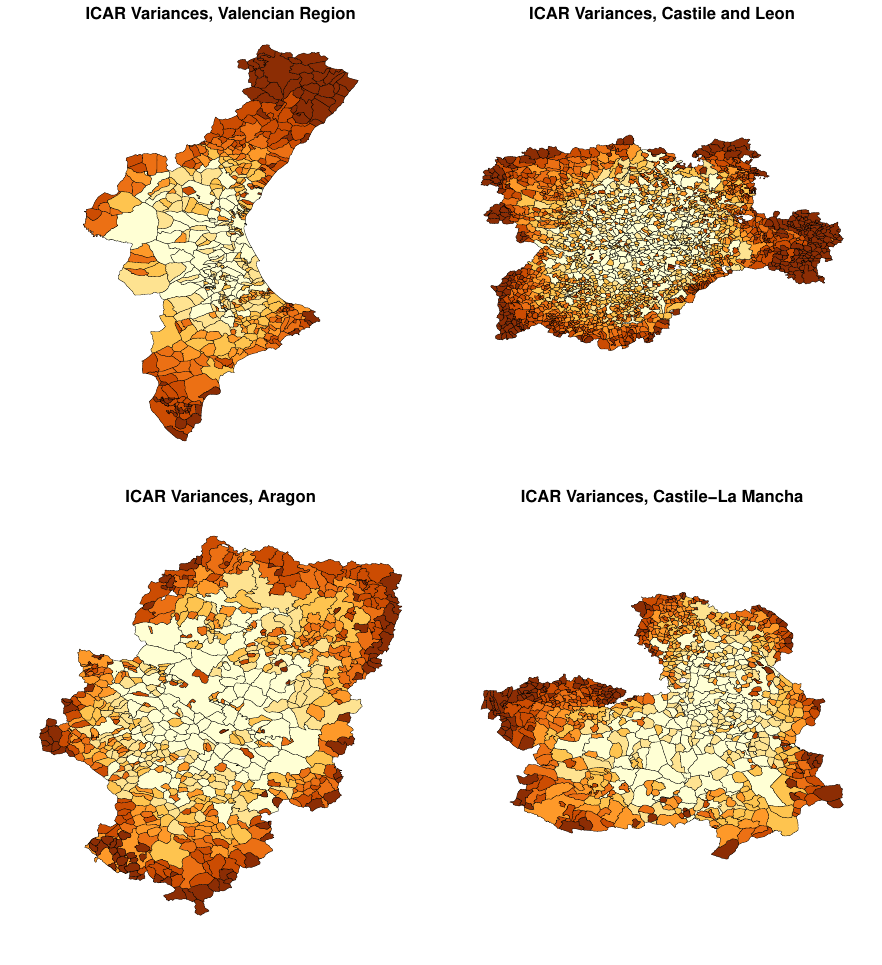}
\centering
\caption{Marginal variance for the municipalities of 4 autonomous regions of Spain with markedly different shapes. Variances correspond to an ICAR distribution over the corresponding region. Darker municipalities stand for higher variances.}
\label{fig:PriorVariances}
\end{figure}

As previously mentioned, the conditional character of CAR distributions, and the ICAR distribution in particular, could yield counter-intuitive features in marginal terms. Therefore, we are going to explore the performance of the marginal variances of the ICAR distribution in some applied settings. In particular, Figure \ref{fig:PriorVariances} shows the marginal variance for each municipality of 4 of the autonomous regions (out of 17) that compound Spain. These regions have been chosen with illustrative purposes according to their different geometries. The marginal variance for the \(i\)-th municipality in each plot has been derived as \(\Sigma_{ii}=(\mathbf{Q}^-)_{ii}\), where \(\mathbf{Q}^-\) stands for the Moore-Penrose pseudoinverse of \(\mathbf{Q}\). A more detailed discussion on this issue is given in the following section. Without loss of generality the precision parameter \(\tau\) has been fixed to 1 for this whole study.

The irregular shapes of the regions in Figure \ref{fig:PriorVariances} could have made us guess that the marginal municipal variances had an unpredictable character. However, those maps show clear geographical patterns whose recurrence points out to a general artifact of the ICAR distribution. First, we notice a clear common edge effect for all of them, which makes the municipalities placed at their borders to show higher variances, in general, than the municipalities in their inner side. This is the most remarkable feature in these maps, nevertheless the maps in Figure \ref{fig:PriorVariances} show some additional common characteristics. For example, variances at the center of each region show much less variability than any other location of the regions of study. However, some municipalities of high variance can be noticed at these regions. Those municipalities typically correspond to areas with a low number of neighbours, usually no more than one or two. This may be understood as a consequence of the conditional character of ICAR distributions, which makes the conditional variance of municipalities with few neighbours to be higher and that high conditional variance could make the corresponding marginal variances to be also high. Additionally, it also seems remarkable the similar performance of the marginal variances for all 4 regions, thus for the Valencian Region the minimum, median and maximum variances are equal to 0.20, 0.56 and 2.02, for Castile and Leon these values are 0.16, 0.49 and 2.07, for Aragon 0.13, 0.49 and 2.00 and for Castile-La Mancha 0.16, 0.54 and 1.94. Therefore, at least for these 4 settings, the location and range of the marginal variances seem to be mostly constant, regardless of the size and shape of the region of study, although this could possibly depend on the number of neighbours of the units of the region of study.

An interesting feature of the geographical distribution of the marginal variances is that it may also depend on the geometric shape of each region of study. For example, for the Valencian Region and Aragon, with long-shaped geographic structures, the municipalities with higher variability are at the north and south of those regions, which correspond to regions with higher eccentricity values (greater distances to the farthest area) in the corresponding graph. Therefore, in those cases, the long-shaped structure of the region of study induces artifacts on the geographical distribution of the marginal variance as clear as the previously mentioned edge effect. On the other hand, both Castile and Leon and Castile-La Mancha are more round-shaped than the Valencian Region and Aragon so we cannot talk about any long axis traversing these regions. However, we can notice another artifacts for both two coming from the geographic shape of those regions. Thus, for Castile and Leon we notice several regions of high variance close to the border but no so on the border. The more evident of these, although not the only one, is at the eastern protrusion of this region, where the isolated character of this region makes it show an increased variance, beyond the edge effect that could show the municipalities at the borders in this region. Similar effects can be noticed for Castile-La Mancha, particularly evident at the western part of this region. Therefore, as a summary, beyond previously known and documented edge effects, the conditional character of ICAR distributions makes additional effects to arise in the marginal variances of the process, related with the particular geometry of each region of study.

\begin{figure}
\centering
\includegraphics[width=0.7\textwidth]{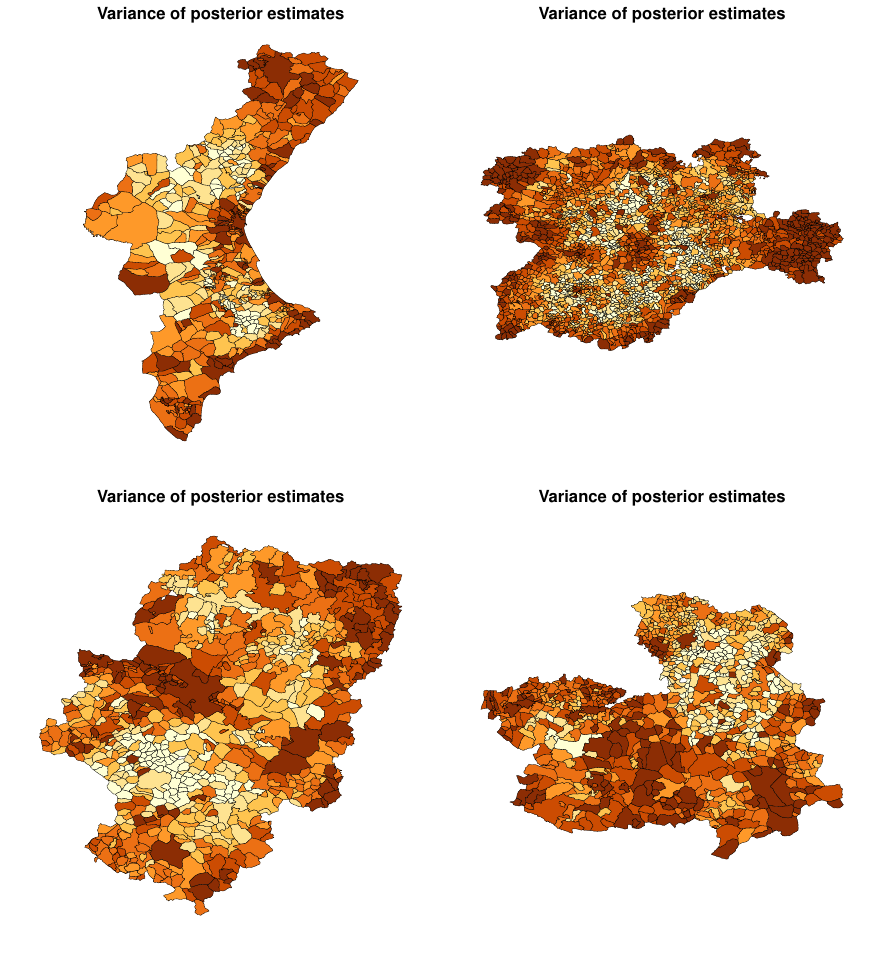}
\centering
\caption{Empirical variances of the posterior mean estimates of the log-risks for each municipality across the 100 causes of death analyzed in each region. Darker municipalities stand for higher variances.}
\label{fig:PosteriorVariances}
\end{figure}

The illustrated effects are intrinsic to the ICAR distribution, however, this distribution is frequently used in hierarchical Bayesian models which include random effects with this distribution to allow for geographical variability, and dependence, of some quantity of interest. Thus, an interesting question arises here: Are the mentioned artifacts of the ICAR distribution noticeable in practice within the results of hierarchical models that incorporate it? To investigate this, we have examined 102 causes of death from the Spanish National Statistical Institute across the four regions depicted in Figure  \ref{fig:PriorVariances} in the period 1989-2014. For these causes, we have selected those with at least 16.000 deaths during that period either for men or women, separately. Since Spain is composed by about 8.000 municipalities this means that our selection contains the causes of death with at least 2 expected deaths per municipality, on average, during the period of study. These yields a collection of 100 causes of death, 50 for men and 50 for women. For each of these causes and regions we have run the corresponding Besag, York and Mollié (BYM) model \citep{Besag.York.ea1991}. More details on the BYM model can be found in the following section. All the models in this paper have been run in \texttt{INLA}. For each municipality, we have calculated the empirical variance of the posterior means of the (log-)linear predictor over the 100 analyses carried out. In this manner we estimate the variability of the posterior mean estimates for each municipality for this large collection of causes of death, which is intended to randomize the geographical spatial patterns that could be found over these regions (see Figure \ref{fig:PosteriorVariances}).

These variances may be conditioned by other aspects beyond the geometry of the region of study. For example, the distribution of more or less populated municipalities over the regions of study show, in general  (plots in Figure \ref{fig:PosteriorVariances}), higher variances in the most populated areas, which are probably due to an increased statistical power in these regions. In addition, the inclusion of the independent random effect in the BYM model, which is not present in the plots shown in Figure \ref{fig:PriorVariances}, could be having also a relevant effect. However, the plots in Figure \ref{fig:PosteriorVariances} also resemble those in Figure \ref{fig:PriorVariances} in different aspects. Particularly, the correlation between the variances represented in both figures is 0.38 for the Valencian Region, 0.52 for Castile and Leon, 0.19 for Aragon, and 0.34 for Castile-La Mancha. Thus, it seems that the heteroscedastic effects that we foresaw for the prior dependence structure of ICAR distributions could be possibly translated in practice, at least in our studies, to the posterior estimates.

\section{A proposal of CAR distribution favouring homoscedasticity}

As described, regular ICAR-distributed random vectors show, by construction, important undesirable features in terms of (prior) marginal variances inherited from their conditional formulation. We are going to propose now a modification of the ICAR distribution intended to fix those effects. This section is structured in two parts. The first part discusses generalized inverses of matrices and their relationship with the ICAR distribution. In this subsection we will introduce some results that will be useful for the subsequent part. The second subsection introduces a proposal of CAR distribution which favours homoscedasticity, which is the main goal of this section.

\subsection{On generalized inverses and their relationship with the ICAR distribution}

ICAR distributions have rank-deficient precision matrices. The dimension of that rank-deficiency coincides with the number of connected components of the graph that summarises the region of study \citep{Hodges.Carlin.ea2003}. For simplicity, as usual, we will consider from now on that such region has a single connected component, although all of the results below could be also derived for regions with more than one connected component, just bearing in mind that the corresponding rank-deficiency of $\mathbf{Q}$ is higher than 1. The vector $\mathbf{1}_I$ will be an eigenvector associated with a null eigenvalue of the structure matrix $\mathbf{Q}$ of any ICAR, since
\[\mathbf{Q}\mathbf{1}_I=\mathbf{D}\mathbf{1}_I-\mathbf{W}\mathbf{1}_I=(\sum_{\{j:j\sim 1\}} w_{1j}-\sum_{\{j:j\sim 1\}} w_{1j},\ldots,\sum_{\{j:j\sim I\}} w_{Ij}-\sum_{\{j:j\sim I\}} w_{Ij})'=\mathbf{0}_I.\]
Therefore, the null space of $\mathbf{Q}$ goes along the direction defined by $\mathbf{1}_I$. In order to fix the rank-deficiency of \(\mathbf{Q}\), ICAR-distributed vectors \(\pmb{\theta}\) are typically constrained along (made orthogonal to) its null space. Therefore, the constrain \(\pmb{\theta}\textbf{1}_I=\sum_i \theta_i=0\) is usually assumed \citep{Goicoa.Adin.ea2018}, which restricts the multivariate vector \(\pmb{\theta}\) to an $(I-1)$-dimensional subspace, that spanned by the non-null eigenvectors of \(\mathbf{Q}\). In this manner, \(\pmb{\theta}\) avoids the direction of the space which makes the ICAR distribution improper and is the restricted to the subspace where the corresponding multivariate normal distribution is proper.



Generalized inverses are closely related to the ICAR distribution. Those distributions are, at the end, multivariate normal distributions with singular precision, and therefore singular covariance matrices. Thus, the density function of that multivariate normal distribution depends on $\mathbf{Q}^{-1}$, that does not exist in practice. As a consequence, generalized inverses of $\mathbf{Q}$ are required in order to define that density function. A generalized inverse of a matrix $\mathbf{Q}$ is any matrix $\mathbf{Q}^g$ with the property $\mathbf{QQ^gQ}=\mathbf{Q}$ \citep{Gentle2007}. Evidently, if $\mathbf{Q}$ was invertible, $\mathbf{Q}^{-1}$ would fulfill this property so, in some way, generalized inverses emulate inverse matrices in some manner when these do not exist.

Generalized inverse matrices of a singular matrix are not unique. Thus, let us assume that $\mathbf{U}\mathbf{D}\mathbf{U}'$ is the eigen-decomposition of $\mathbf{Q}$, with
$$\mathbf{D}=\left(
\begin{matrix}
\mathbf{D}_{I-1} & \mathbf{0}_{I-1} \\
\mathbf{0}_{I-1}' & 0
\end{matrix}
\right)$$
and $\mathbf{D}_k=diag(d_1,\ldots,d_{I-1})$, where $\{d_i,1=1,\ldots,I-1\}$ are the non-null eigenvalues of $\mathbf{Q}$. Then, $\mathbf{Q}^g$ is a generalized inverse of $\mathbf{Q}$ if and only if $\mathbf{Q}^g=\mathbf{U}\mathbf{D}^*\mathbf{U}'$, where
\begin{equation}
\mathbf{D}^*=\left(
\begin{matrix}
\mathbf{D}_{I-1}^{-1} & \mathbf{x} \\
\mathbf{y}' & z
\end{matrix}
\right)
\label{Dstar}
\end{equation}
for vectors $\mathbf{x},\mathbf{y}$ of length $I-1$ and a scalar $z$ \citep{Harville1997}. For a multivariate normal vector $\pmb{\theta}$ of length $k$, with mean $\pmb{\mu}$ and singular covariance matrix $\mathbf{\Sigma}$, the density function may be written as:
$$P(\pmb{\theta}\mid\pmb{\mu},\pmb{\Sigma})=\frac{(2\pi)^{k/2}}{\sqrt{d_1\cdot\ldots\cdot d_{I-1}}}\exp\left(-(\pmb{\theta}-\pmb{\mu})'\pmb{\Sigma}^g(\pmb{\theta}-\pmb{\mu})\right)$$
for any choice of the generalized inverse $\pmb{\Sigma}^g$ (\cite{Rao2002}, page 528). Therefore, that density function is invariant to the choice of the generalized inverse of the corresponding (singular) covariance matrix.

Regretfully, some other features of singular multivariate normal distributions do depend on the choice of the corresponding generalized inverse. For example, for an ICAR distribution with structure matrix $\mathbf{Q}$, the corresponding covariance matrix does not match with every generalized inverse of $\mathbf{Q}$ since that covariance matrix is unique while $\mathbf{Q}^g$ takes many different (and not necessarily symmetric) forms. Therefore, only one of the generalized inverses of $\mathbf{Q}$ will coincide with the covariance matrix of $\pmb{\theta}$. Since that covariance matrix is symmetric, it is evident that the generalized inverse of $\mathbf{Q}$ leading to the covariance matrix should have a symmetric $\mathbf{D}^*$ matrix in expression (\ref{Dstar}), so the generalized inverse corresponding to the covariance matrix should have $\mathbf{x}=\mathbf{y}$. On the other hand, bearing in mind that for an ICAR distribution the mean of $\pmb{\theta}$ is $\mathbf{0}_I$, if $\pmb{\Sigma}$ was the corresponding covariance matrix, then:
\[(\pmb{\Sigma}\cdot\mathbf{1}_I)_i=\pmb{\Sigma}_{i\cdot}\cdot\mathbf{1}_I=E(\theta_i^2)+\sum_{j=1}^I E(\theta_i\theta_j)=E\left(\theta_i\sum_{j=1}^I\theta_j\right)\]
which is equal to 0 if $\sum_{j=1}^I \theta_j=0$, as typically assumed. Thus, if this constrain is imposed, the generalized inverse $\mathbf{Q}^-$ coinciding with the covariance matrix should fulfill
$$\mathbf{0}_I=\mathbf{Q}^-\mathbf{1}_I=\mathbf{U}\mathbf{D}^*\mathbf{U}'\mathbf{1}_I=\mathbf{U}\mathbf{D}^*(0,\ldots,0,1)'=\mathbf{U}(\mathbf{x}',z)'.$$
As a consequence, $(\mathbf{x}',z)'$ is orthogonal to all the columns of $\mathbf{U}$, which is an orthonormal basis of $\mathcal{R}^I$. This may only happen if $(\mathbf{x}',z)'=\mathbf{0}_I$. Therefore, since both the symmetry and the sum-to-zero conditions have to be fulfilled, $\mathbf{D}^*$ will have to take, necessarily, the form
$$\mathbf{D}^-=\left(
\begin{matrix}
\mathbf{D}_k^{-1} & \mathbf{0}_{I-1} \\
\mathbf{0}'_{I-1} & 0
\end{matrix}
\right).$$
In that case, the generalized inverse yielding the covariance matrix of $\pmb{\theta}$ will be $\mathbf{Q}^-=\mathbf{U}\mathbf{D}^-\mathbf{U}'$, which is the well-known Moore-Penrose generalized inverse (or pseudoinverse) of $\mathbf{Q}$ \citep{Harville1997}. Henceforth, we will refer to the Moore-Penrose pseudoinverse of any matrix by adding the superindex $-$ to the corresponding matrix.

An interesting property of the Moore-Penrose pseudoinverse is that it is consistent with respect to orthogonal transformations. That is, if $\mathbf{Q}$ is a singular structure matrix and $\mathbf{V}$ is an orthogonal matrix, then
$$(\mathbf{V}\mathbf{Q}\mathbf{V}')^-=\mathbf{V}\mathbf{Q}^-\mathbf{V}'.$$
Regretfully this does not happen for non-orthogonal transformations, not even for simple diagonal transformations \citep{Uhlmann2018}. That is, for a general diagonal matrix $\pmb{\Lambda}$, in general
$$(\pmb{\Lambda}\mathbf{Q}\pmb{\Lambda})^-\neq \pmb{\Lambda}\mathbf{Q}^-\pmb{\Lambda}.$$
These properties will be of interest soon below.

\subsection{A CAR distribution that promotes homoscedasticity}

Let us assume that \(\pmb{\theta}\) follows an ICAR distribution with variance-covariance matrix \(\pmb{\Sigma}=(\tau\mathbf{Q})^-\), given by the Moore-Penrose pseudo-inverse of its precision matrix. Let us also denote \(\pmb{\Lambda}=diag(\sigma_1^2,\ldots,\sigma_I^2)\) the diagonal matrix whose diagonal elements coincide with those of \(\pmb{\Sigma}\), the marginal variances of every spatial unit. In that case, let us consider now the modified ICAR distribution defined by \(N_I(\mathbf{0}_I,\tau(\pmb{\Lambda}^{1/2}\mathbf{Q}\pmb{\Lambda}^{1/2}))\). This distribution has also a sparse precision matrix, as sparse as the original matrix \(\mathbf{Q}\) since they will share the same zero cells. The difference between \(\mathbf{Q}^*=\pmb{\Lambda}^{1/2}\mathbf{Q}\pmb{\Lambda}^{1/2}\) and
\(\mathbf{Q}\) will be that the non-zero cells of \(\mathbf{Q}^*\) will be weighed but those of \(\mathbf{Q}\), generally, will be not (although if they were, \(\mathbf{Q}^*\) could be similarly defined as a reweighed version of \(\mathbf{Q}\)). If $\mathbf{Q}$ was not singular, then $(\mathbf{Q}^*)^-=(\mathbf{Q}^*)^{-1}=\tau^{-1}(\pmb{\Lambda}^{-1/2}\mathbf{Q}^{-1}\pmb{\Lambda}^{-1/2})=\pmb{\Lambda}^{-1/2}\pmb{\Sigma}\pmb{\Lambda}^{-1/2}$ would have all its diagonal terms equal to one, what would make this new modified CAR distribution homoscedastic. Regretfully, for ICAR distributions $\mathbf{Q}$ will be singular and, in that case, as mentioned at the end of the previous subsection,
$(\mathbf{Q}^*)^-=(\pmb{\Lambda}^{1/2}\mathbf{Q}\pmb{\Lambda}^{1/2})^-\neq\pmb{\Lambda}^{-1/2}\mathbf{Q}^-\pmb{\Lambda}^{-1/2}$. Therefore, our proposed transformation, in contrast to the non-singular case, would not necessarily lead to a pure homoscedastic process. However, this transformation should enhance homoscedasticity since the conditional variance is decreased(/increased) wherever the original ICAR showed particularly high(/low) marginal variances. This should balance the initially more heterogeneous variances of the ICAR distribution, leading to more similar values. Therefore, we will call from now on this new CAR distribution as \emph{HomCAR}, as an abbreviation of \emph{Homoscedastic CAR}, although the homoscedastic term here should be intended in a broad, not so strict, manner.

From a conditional point of view, the HomCAR distribution could be alternatively formulated as the following set of conditional distributions:
\begin{equation}\theta_i\mid \pmb{\theta}_{-i}\sim N\left(-\sum_{\{j:j\sim i\}}\frac{Q_{ij}^*}{Q_{ii}^*}\theta_j,\tau Q_{ii}^*\right)=N\left(n_i^{-1}\sum_{\{j:j\sim i\}}\frac{\sigma_j}{\sigma_i}\theta_j,\tau (n_i\sigma_i^2)\right),\;i=1,\ldots,I.\label{conditional}\end{equation}
In this expression, the conditional expected value of each observation is the mean of some rescaled version of the neighbouring values. The rescaling factor $\sigma_j/\sigma_i$ overweighs(/underweighs) the contribution of the neighbouring locations with high(/low) marginal variance. This makes the high variance of the observations with that feature to be transferred to their neighbouring sites, balancing the variances between those locations. In contrast to regular ICARs, the conditional means in (\ref{conditional}) are no longer weighted means of the neighbours' values $\theta_j$, since the weights in the new linear combinations ($n_i^{-1}\sigma_j/\sigma_i$) do not add up to 1. In relation to this issue, if a specific region $i$ exhibits particularly high (or low) marginal variance $\sigma_i$ in the original ICAR model, the weights in the linear combination used to compute its mean in the HomCAR distribution will be correspondingly low(or high). This adjustment reduces(or increases) the variability of this region, thereby achieving a leveling effect. Regarding the conditional precisions in expression (\ref{conditional}), regions with higher (or lower) marginal variance in the original model will now exhibit higher (or lower) conditional precision. This increased conditional precision causes $\theta_i$ to align more closely with its neighbours, further reducing its marginal variance.

The HomCAR distribution has close similarities to the CAR distribution proposed by \cite{Cressie.Chan1989} and \cite{Stern.Cressie1999}.
This CAR distribution has a structure matrix given by:
\begin{equation}
Q_{ij}=\left\{ \begin{aligned}
 E_i & &\textrm{if } i=j,\\
 -\phi (E_iE_j)^{1/2} & &\textrm{if } i\sim j,\\
 0 & &\textrm{otherwise}.
\end{aligned} \right.
\end{equation}
where $\{E_i,\;i=1,\ldots,I\}$ stands for the expected values associated to each geographical location. The presence of the parameter $\phi$ in this distribution makes it a kind of proper CAR distribution instead of a modification of the ICAR. However, leaving aside this parameter, the intrinsic version ($\phi=1$) of this distribution would be equivalent to the HomCAR with the $E_i$ replaced by $\sigma_i^2$. The rationale for the original formulation of this CAR distribution is that, for any two neighbouring sites $i$ and $j$, the partial correlation coefficient $corr(\theta_i,\theta_j\mid \pmb{\theta}_{-(i,j)})$ is constant. This does not happen for the traditional ICAR with all their weights equal to 1, where those partial correlations are equal to $(n_in_j)^{-1/2}$ \citep{Stern.Cressie1999}, that is, they depend on the number of neighbours. That constant partial correlation seems, and is perceived as, a good property by Stern and Cressie. Interestingly, that property is also shared with the HomCAR distribution, since for a proper formulation of this distribution the partial correlation \citep{Stern.Cressie1999}
would be $\phi(\sigma_i/\sigma_j)(\sigma_j/\sigma_i)=\phi$. Thus, the HomCAR, beyond favouring homoscedasticity, seems to show also benefits in terms of constant partial correlations between neighbouring sites.


The rank-deficiency of the HomCAR distribution has the same dimension than that of its original ICAR distribution. However, although rank-deficient, the HomCAR cannot be seen as a particular case of a traditional ICAR in the sense that \(\mathbf{Q}^*\) cannot be expressed in the form \(\mathbf{D}^*-\mathbf{W}^*\), where the diagonal elements \(D^*_{ii}\) are equal to \(\sum_{j\sim i}W_{ij}\). For the HomCAR, the sum of the cells of each row of \(\mathbf{Q}^*\) are equal to \(\sigma^2_in_i-\sigma_i\sum_{\{j:j\sim i\}}\sigma_j,\;i=1,\ldots,I\), which are generally different from zero, in contrast to the ICAR.

The rank-deficiency of \(\mathbf{Q}^*\) is inherited from \(\mathbf{Q}\) and, then, they share the same number of null eigenvalues. However, the corresponding null eigenvectors will not be the same for these two matrices. Thus, if \(\mathbf{Q}=\mathbf{U}\mathbf{D}\mathbf{U}'\) let us define $\mathbf{u}_I^*=\alpha\pmb{\Lambda}^{-1/2}\mathbf{U}_{.I}=\alpha\pmb{\Lambda}^{-1/2}\mathbf{1}_{I}$, for $\alpha$ a constant which makes the norm of $\mathbf{u}_I^*$ equal to 1. Therefore,
\[\mathbf{Q}^*(\mathbf{u}_I^*)=(\mathbf{\Lambda}^{1/2}\mathbf{Q}\mathbf{\Lambda}^{1/2})(\alpha\mathbf{\Lambda}^{-1/2}\mathbf{1}_I)=\alpha\mathbf{\Lambda}^{1/2}\mathbf{Q}\mathbf{1}_I=\mathbf{0}_I,\]
that is, if \(I^{-1/2}\mathbf{1}_I\) is the eigenvector associated to the null eigenvalue of \(\mathbf{Q}\), \(\mathbf{u}^*_I\) will be the corresponding eigenvector of \(\mathbf{Q}^*\) associated to that same eigenvalue. Thus, the direction along which the HomCAR distribution is improper is different to that of the ICAR distribution. 

Since $\mathbf{u}_I^*$ is proportional to \(\mathbf{\Lambda}^{1/2}\mathbf{1}_I\), instead of \(\mathbf{1}_I\), the confounding between the intercept and the HomCAR distribution will be minor than for the ICAR. However, it would still be desirable to impose some constrain over the HomCAR in order to make it proper on the constraint domain. In that case, the suitable restriction would not be a sum-to-zero constrain, but a restriction of the kind \(\pmb{\theta}'\mathbf{\Lambda}^{1/2}\mathbf{1}_I=\sum_{i=1}^I \theta_i\sigma_i=0\) in order to make $\pmb{\theta}$ orthogonal to the direction which defines the impropriety of the HomCAR distribution. 

ICAR-distributed random effects are frequently used in hierarchical Bayesian models through the BYM model \citep{Besag.York.ea1991}. For count data \(O_i\) associated to a series of \(I\) spatial units, this model is usually formulated as:
\[O_i\sim Poisson(E_i\cdot\mu_i),\;i=1,\ldots,I\]
\[\log(\mu_i)=\beta_0+u_i+v_i\]
where the vector \(\mathbf{u}\) follows an ICAR distribution of precision parameter \(\tau_u\) and associated structure matrix \(\mathbf{Q}\) and \(v_i\sim N(0,\tau_v)\) take independent normal values. In this model, the spatially structured random effect \(\mathbf{u}\) could be replaced by a new spatial random effect with structure matrix \(\mathbf{Q}^*=\pmb{\Lambda}^{1/2}\mathbf{Q}\pmb{\Lambda}^{1/2}\), as above. In that case, the covariance matrix of
\(\mathbf{u}+\mathbf{v}\), which defines the geographical variability of the log-risks would be \(\tau_u^{-1}(\mathbf{Q}^*)^-+\tau_v^{-1}\cdot\mathbf{I}_I\) which, as described, should be closer to homoscedasticity than the covariance matrix \(\tau_u^{-1}\mathbf{Q}^-+\tau_v^{-1}\cdot\mathbf{I}_I\) of \(\mathbf{u}+\mathbf{v}\) in the traditional BYM model.

\section{A practical assessment of the HomCAR distribution}

In this section we are going to assess in practice the performance of the HomCAR distribution introduced in the previous section. First, we compare the (prior) marginal variances of the HomCAR distribution with those of the ICAR for each of the 4 regions of study considered in Section 2. The minimum, median and maximum municipal variances are 0.81, 0.98 and 1.13 for the Valencian Region, 0.93, 1 and 1.06 for Castile and Leon, 0.92, 1 and 1.05 for Aragon and 0.87, 0.99 and 1.09 for Castile-La Mancha. Once again these values are very similar for all 4 regions and show some variability that illustrates how the HomCAR distribution, as previously commented, is not purely homoscedastic. However, the ranges for the marginal variances for the HomCAR distribution are far lower than for the original ICARs. For the ICARs the (prior) maximum marginal variance is always, at least, 10 times higher than the minimum, in contrast to the HomCAR where that relationship is always lower to 1.4. This illustrates the improvement in homoscedasticity achieved by the HomCAR distribution.

We will explore now in practice the performance of the HomCAR distribution, focusing on the corresponding empirical posterior variances of the risks instead of their prior variances. First, we will do it over a collection of simulated data sets and finally we will revisit the analysis of the Spanish mortality data sets (Section 2), now using the HomCAR distribution.

\subsection{Performance on simulated data sets}

For the assessment of our proposal we have generated the following synthetic data sets: For each of the regions of study in Section 2 we have simulated 100 different data sets. The underlying risk patterns for all of them have been assumed to follow a (continuous) Gaussian field of mean 0 and variance-covariance matrix given by
\[Cov(\theta_i,\theta_j)=(0.2)^2\exp\left(-3\frac{d_{ij}}{\delta}\right),\]
where \(d_{ij}\) denotes the distance between the centroids of spatial units \(i\) and \(j\). This produces a homoscedastic covariance function of marginal standard deviation equal to 0.2 for every municipality. The parameter \(\delta\) will change for each region and is chosen as the 0.25 quantile of the distances between spatial units for that region, therefore those municipalities separated by that distance will show a correlation of \(\exp(-3)\approx 0.05\), so \(\delta\) may be understood as the spatial range of the process. As mentioned, we have generated 100 different spatial patterns, \(\pmb{\theta}\), for each region according to this process and, for each of them, we have generated a sample \(\mathbf{O}\) of observed cases from a Poisson distribution of mean \(5\cdot \exp(\pmb{\theta})\). In this manner we have considered a constant number of expected cases per municipality in order to avoid the effect of that variability on the corresponding vectors of observed cases \(\mathbf{O}\).

For each of these simulated data sets we have run two models, first, the traditional BYM model and, second, an alternative `BYM' model with the original ICAR replaced by a HomCAR-distributed random effect. This second model is not a genuine BYM model since the HomCAR distribution is not really an ICAR, however they share the same aim as both two split the variability between spatially structured an unstructured terms. It has been necessary to resort to the \texttt{generic0} latent family of \texttt{INLA} to implement the HomCAR as the \texttt{besag}, \texttt{bym} and \texttt{bym2} options define themselves the diagonal of the structure matrix and our proposal requires our own particular definition. 
A Rmarkdown (and corresponding pdf) document with the code, containing the full details of the analyses in this section and Section 2, is supplied as supplementary material to the paper.

\begin{figure}
\centering
\includegraphics{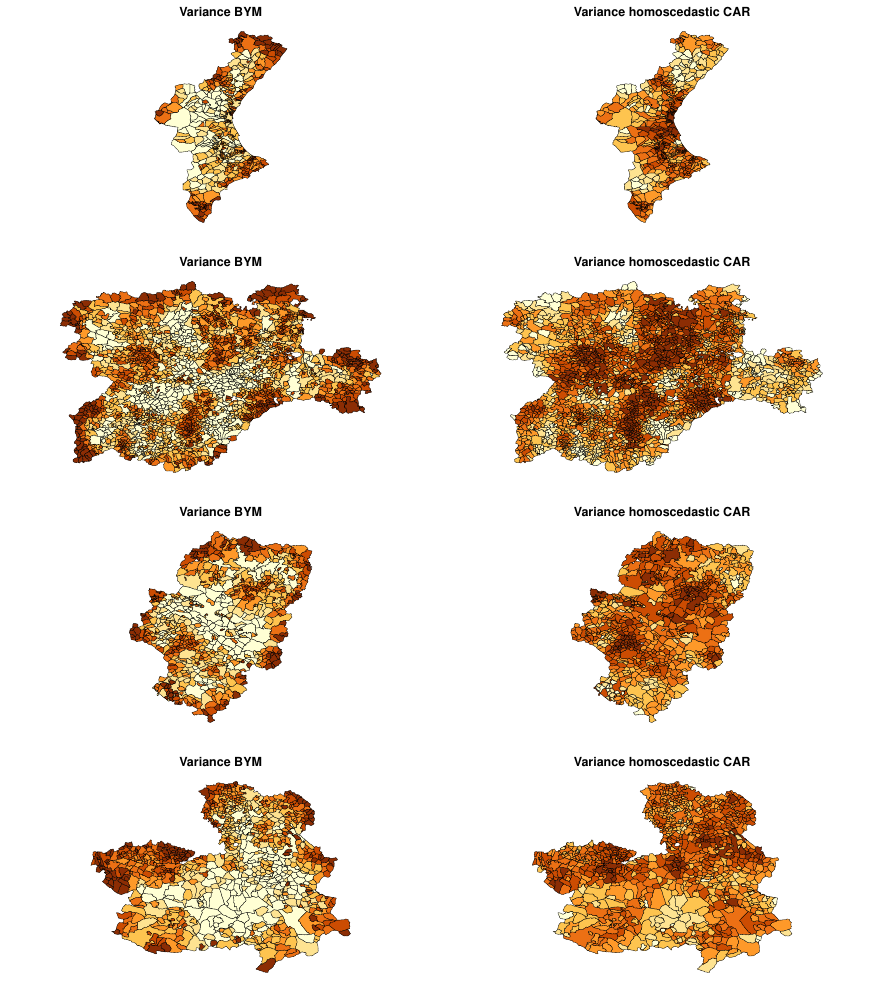}
\caption{Empirical variances of posterior mean estimates for the log-risks across the 100 simulated data sets. Left, variances for the traditional BYM model, and right, variances for that including the HomCAR distribution. Darker municipalities stand for higher variances.}
\label{fig:VariancesSimulated}
\end{figure}

\begin{figure}
\centering
\includegraphics[width=0.7\textwidth]{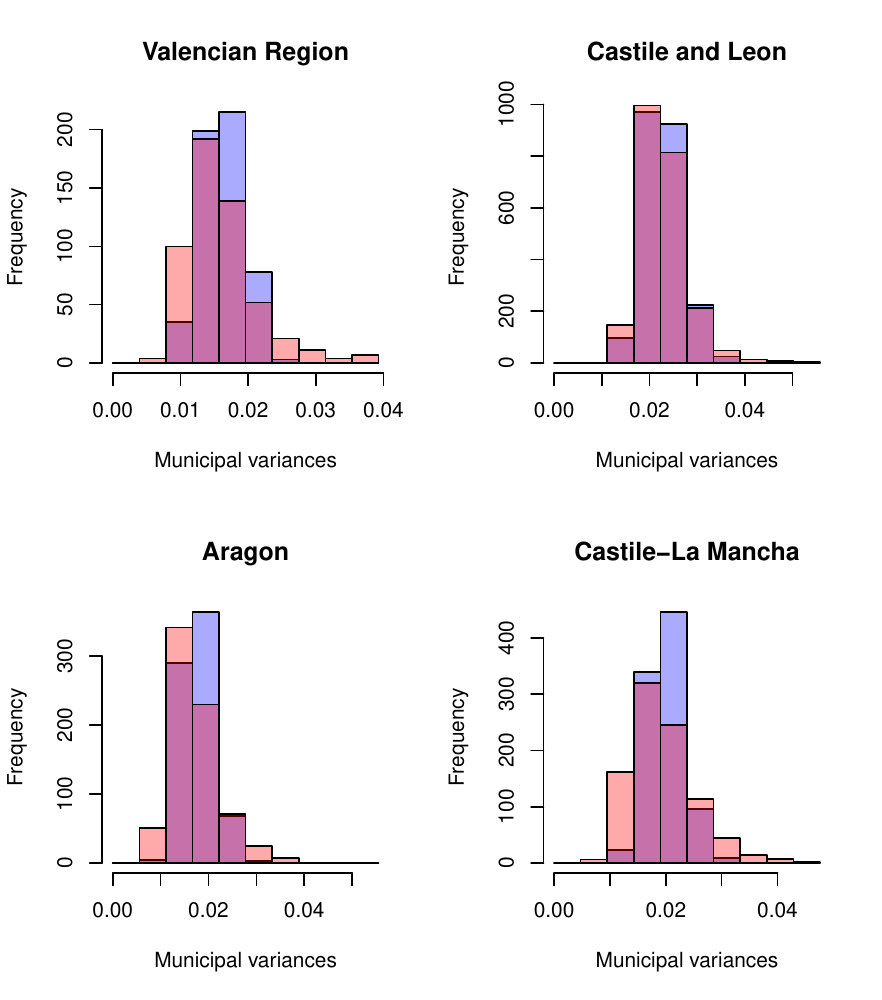}
\centering
\caption{Histograms displaying the municipal empirical variances of posterior mean estimates. Red corresponds to BYM and blue to the homoscedastic model.}
\label{fig:Histograms}
\end{figure}

Figure \ref{fig:VariancesSimulated} displays, for each region, the municipal variances for the 100 estimated (log-)risks for both models. The left column corresponds to the variances of the BYM model and that on the right corresponds to the HomCAR-based model. For comparability, the cuts for both maps in each row are given by the septiles of the variances for the BYM model, although a more detailed comparison of the distribution of those variances for both models will be discussed in Figure \ref{fig:Histograms}. The municipal variances for both models show some random variability corresponding to the particular samples used in the study. However, beyond that random variability, the BYM model shows, systematically, an increased variability in those areas which also showed increased variance in Figure \ref{fig:PriorVariances}. As a consequence, the results of the BYM model show also edge effects and geometric features that we would not want to reproduce. On the contrary, we do not notice those effects in the results of the HomCAR-based model which show, supposedly random, variability that we cannot attribute to any known feature of the regions of study. Interestingly, the geographical patterns reproduced by the HomCAR model can also be noticed in the results of the BYM model, although those underlying patterns are mostly hidden by the edge and geometric artifacts reproduced by the ICAR distribution which induce higher variance at some places. As a consequence, those features are the more prominent feature (or artifact) reproduced by the BYM model in this study. Regretfully, these artifacts will be generally unnoticed when isolated (univariate) maps, corresponding to the study of a single disease, are typically carried out.

Figure \ref{fig:Histograms} shows for each region a histogram for the empirical variances of the municipal estimates of both models. Red and blue columns correspond, respectively, to the BYM and HomCAR models. It can be noticed how the HomCAR-based estimates, with more peaked distributions, show, systematically, less variability than BYM. This supports the supposedly homoscedastic nature of this proposal, as it avoids introducing additional artifacts that could inflate that variability among municipalities. In addition, the reduction of variability in the HomCAR histograms come mainly from the reduction of the right tail of the histograms in comparison to those of BYM. Possibly, that right-tail shortening is due to the edge and geometric artifacts that the homoscedastic proposals seems to fix. Figure \ref{fig:Histograms} shows also an unpleasant feature for both models and possibly of smoothing models in general. Thus, the theoretical marginal variance for every municipality is 0.04 and, as shown in the histograms, the fitted variances are systematically lower. This seems a consequence of some oversmoothing that these models, intended to smooth risks, are doing in practice. However, the Mean Squared Error ($I^{-1}\sum (var(\theta_i)-0.04)^2$) for these variances are systematically lower for the HomCAR than for the BYM for all 4 settings. In addition, the mean municipal variance ($I^{-1}\sum var(\theta_i)$) for the HomCAR is also higher than that for BYM in all 4 cases (MSEs and means shown at the annex material). Thus, BYM seems to oversmooth the risks in a larger amount than its homoscedastic counterpart.

\begin{table}[!ht]
\caption{Assessment of model selection criteria and log-risk estimation accuracy.}
\label{tab:Statistics}
\begin{center}
\begin{tabular}{|r|ccccc|}
\hline
{\bf Valencian Region} & DIC & WAIC &  MAB & RMSE & IS \\[0.5ex]
\hline
BYM    & 2392.6 & 2391.1 & 0.0162 & 0.1555 & 0.7746 \\
HomCAR-based  & 2391.5 & 2390.3 & 0.0132 & 0.1543 & 0.7714 \\
\hline
{\bf Castile and Leon} & DIC & WAIC & MAB & RMSE & IS \\[0.5ex]
\hline
BYM    & 10044.1 & 10039.0 & 0.0094 & 0.1281 & 0.6449 \\
HomCAR-based  & 10042.1 & 10038.3 & 0.0110 & 0.1275 & 0.6473 \\
\hline
{\bf Aragon} & DIC & WAIC &  MAB & RMSE & IS \\[0.5ex]
\hline
BYM    & 3292.2 & 3289.7 & 0.0123 & 0.1500 & 0.7416 \\
HomCAR-based  & 3290.4 & 3288.7 & 0.0121 & 0.1485 & 0.7351 \\
\hline
{\bf Castile-La Mancha} & DIC & WAIC & MAB & RMSE & IS \\[0.5ex]
\hline
BYM    & 4126.0 & 4123.8 & 0.0113 & 0.1428 & 0.7070 \\
HomCAR-based  & 4124.2 & 4122.5 & 0.0111 & 0.1416 & 0.7006 \\
\hline
\end{tabular}
\end{center}
\end{table}

Table \ref{tab:Statistics} provides a general comparison of both models, focusing on regions where geographical and edge effects are less likely to occur. That table shows, for each region and model, several model selection criteria along with summary statistics that assess the accuracy of the corresponding (log-)risk estimates. In particular, it contains two model comparison criteria: the Deviance Information Criterion (DIC) and Watanabe-Akaike Information Criterion (WAIC). In addition, 3 statistics assessing the estimation accuracy are shown: the Mean Absolute Bias (MAB) of the posterior mean of the log-risks, their Root Mean Square Error (RMSE) and the Interval Score (IS) \citep{Gneiting.Raftery2007}. The MAB for the whole simulated data sets is defined as $MAB=(I\cdot J)^{-1}\sum_{i=1}^I\left|\sum_{j=1}^J \widehat{LR}_{ij}-LR_{ij}\right|$, where $\widehat{LR}_{ij}$ stands for the estimated log-risk for the $i$-th municipality and $j$-th simulated data set and $LR_{ij}$ is the true simulated value. The RMSE is defined as $RMSE=\sqrt{(I\cdot J)^{-1}\sum_{i=1}^I \sum_{j=1}^J (\widehat{LR}_{ij}-LR_{ij})^2}$. The interval score summarizes the width of the prediction intervals for each model and penalizes those intervals that fail to include the true value. Lower values of these criteria and statistics indicate better model performance. In Table \ref{tab:Statistics}, no relevant differences in bias (MAB) are observed between the two models. However, small but consistent differences in DIC, WAIC, RMSE and IS favour the HomCAR-based model, suggesting a better overall fit for this model. These results reinforce the previously shown homoscedastic advantages of the HomCAR-based model for these simulated data sets.

\subsection{Homoscedastic analysis of the Spanish mortality data sets}

\begin{figure}
\centering
\includegraphics[width=0.7\textwidth]{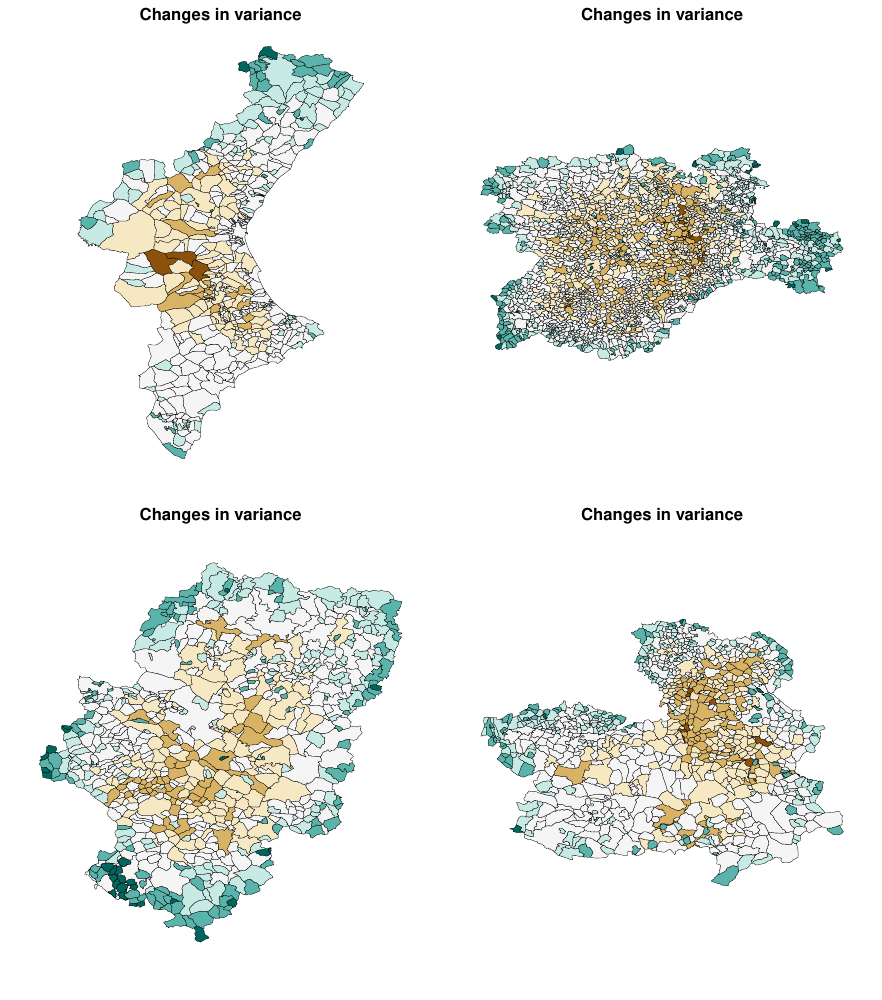}
\centering
\caption{Differences in variance of the relative risks between HomCAR-based and ICAR-based BYM models. Brown areas correspond to locations with higher variance for the HomCAR model while green areas show lower variance for this model.}
\label{fig:DifVariances}
\end{figure}

We reanalyze now, with the homoscedastic model, the Spanish mortality data sets already studied in Section 2. We have run that model for the 100 mortality data sets for each region and compared the results with those of the original BYM model. Figure \ref{fig:DifVariances} shows the differences (variance for the HomCAR model less that of the BYM model) between the municipal empirical variances of both models. These differences are expressed in relative terms by dividing them by the corresponding municipal variance of the BYM model. Brown municipalities correspond to those where the municipal variance has increased and green areas correspond to those where it has decreased. White areas have changes in the variances lower than 10\% and darker colours correspond, in this order, to changes from 10\% to 25\%, from 25\% to 50\% and from 50\% to 100\%.

As it can be appreciated in Figure \ref{fig:DifVariances}, variances may increase or decrease even more than a 50\% from one model to the other, illustrating how the HomCAR model reduces the artifacts evidenced by the ICAR distribution in Section 2. Specifically, the empirical variance is increased in the center of the regions of study and decreased in their borders and more isolated areas. In fact, the empirical variances found for the HomCAR model do not longer correlate with the prior ICAR variances since those correlations are now equal to 0.05 for the Valencian Region, 0.12 for Castile and Leon, -0.15 for Aragon and 0.02 for Castile-La Mancha.

\section{Concluding remarks}

Disease mapping, as a research area, has as main goal the control of the variance of the risk estimates, mainly for the smaller areas, so that they allow to estimate the underlying geographical risk patterns in a reliable manner. However, little attention has been previously paid to the marginal variance artifacts that traditional disease mapping modeling tools, such as ICAR distributions, could induce on the estimated patterns. This work sheds light on that issue, unveiling some artifacts that, by definition, should be present in ICAR-based models and checking that, in practice, those artifacts remain in the analysis of real mortality data. Regretfully, those artifacts are very hard, if even possible, to detect from isolated (univariate) studies so their effects could be easily spread throughout a large number of studies in the literature. Fortunately, this work proposes a manner to fix the edge and geometric effects unveiled throughout this paper which, as shown in this paper, may be more significant and widespread in the outcomes of disease mapping studies than previously anticipated.

According to the results in this paper, one could perhaps conclude that the use of CAR distributions for disease mapping, or for the analysis of areal data in general, should be discouraged in comparison to the use of (continuous) Gaussian fields. In our opinion this is not necessarily true. Gaussian fields show the advantage of being homoscedastic unless different marginal variances are explicitly considered in the corresponding model. In fact, since they induce spatial dependence directly over the covariance matrix, they should have very good properties in marginal terms. However, since they are defined under a marginal perspective, their performance will be not possibly so good under a conditional point of view. For example, for the simulated setting considered (Subsection 4.1) in the Valencian Region, the conditional precision of one municipality given the rest may be up to 15.8 larger than in others (20.0 for Castile and Leon, 10.3 for Aragon or 12.4 for Castile-La Mancha). This is a function, mainly, of the proximity of its neighbours, which depends mostly on their geographical size. In general, we should not want geographically smaller units to have different conditional variances, so this is an unpleasant conditional effect of Gaussian random fields that we would prefer not to have.

Along this work we have found that the weighting of the structure matrix of ICAR distributions could be an appropriate solution to their heteroscedastic problems. This procedure has important connections with that in \cite{CorpasBurgos.MartinezBeneito2020} where a weighting of the structure matrix was also sought but in that case for the Leroux et al.'s CAR distribution \citep{Leroux.Lei.ea1999}. They also reached to a similar weighting scheme of the structure matrix, where \(w_{ij}=c_ic_j\) if \(i\sim j\), and 0 otherwise. In that case \(\{c_i,\;i=1,\ldots,I\}\) were parameters of the model that were estimated by following a multivariate approach where several diseases were jointly studied and those parameters were estimated as a function of the common features estimated by those diseases. They found that, in accordance to our results, high values for \(c_i\) make \(w_{ij}\) to be high for \(j\sim i\) and this makes the random effect \(\theta_i\) to be closer to those of their neighbours. This is similar to what we find, for example, to the municipalities in the border of the regions of study, which showed high marginal variances and therefore the corresponding weights \(w_{ij}\) take high values. Interestingly, the authors find also in their work that areas with large \(c_i\) values are often placed at the border of the region of study: `Note that several of these units are located in spatial units at the borders of the graph where the geometry of the graph would impose lesser spatial dependence' \citep{CorpasBurgos.MartinezBeneito2020}. Therefore, both works agree for different reasons in putting higher weights in the more eccentric municipalities of the regions of study.

As a final future line of research of this work, it remains how to adapt this homoscedastic proposal for other CAR distributions such as, for example, the Proper or Leroux et al.'s CAR distribution \citep{Martinez-Beneito.BotellaRocamora2019}. For these distributions the matrices with the marginal variances $\pmb{\Lambda}$ depend on their spatial parameters so they are not fixed, in contrast to the ICAR case worked in this paper. Thus, the proposal of homoscedastic alternatives to these distributions entails additional problems that remain to be solved. This work expects to be a first step on that direction.

\bibliographystyle{apalike}
\bibliography{BibliografiaUTF8}

\end{document}